%% file: sample-sigconf.tex
\documentclass[10pt,nonacm, sigconf]{acmart}
\usepackage[compact]{titlesec} 
\titlespacing{\section}{0pt}{0pt}{0pt}
\usepackage{booktabs}
\usepackage{multirow}
\usepackage{xspace}
\usepackage{graphicx}
\usepackage{epstopdf}
\usepackage{url}
\usepackage{color}
\definecolor{Orange}{rgb}{0.9,0.5,0}
\definecolor{NavyBlue}{rgb}{0.1, 0.4, 0.8}
\definecolor{Magenta}{rgb}{0.8, 0.1, 0.6}
\definecolor{Green}{rgb}{0.1, 0.8, 0.3}
\definecolor{DarkGreen}{rgb}{0.0, 0.7, 0.2}
\definecolor{Brown}{rgb}{0.4, 0.3, 0.1}
\definecolor{Burgundy}{rgb}{0.5, 0.0, 0.13}
\definecolor{BrightCerulean}{rgb}{0.11, 0.67, 0.84}
\definecolor{BlueViolet}{rgb}{0.33,0.1,0.5}

\AtBeginDocument{%
  \providecommand\BibTeX{{%
    \normalfont B\kern-0.5em{\scshape i\kern-0.25em b}\kern-0.8em\TeX}}}
\copyrightyear{2019}
\acmYear{2019}
\setcopyright{rightsretained}
\acmConference{SIGGRAPH '19 Talks}{July 28 - August 01, 2019}{Los Angeles, CA, USA}
\acmDOI{10.1145/3306307.3328180}
\acmISBN{978-1-4503-6317-4/19/07}
\acmBooktitle{SIGGRAPH '19 Talks, 
  July 28 - August 01, 2019, Los Angeles, CA}
\begin{document}
\acmConference[]{}
\acmBooktitle{}
\acmPrice{}
\acmISBN{}
\title{Emotionless: Privacy-Preserving Speech Analysis for Voice Assistants}
\author{Ranya Aloufi, Hamed Haddadi, David Boyle}
\affiliation{%
  \institution{Systems and Algorithms Laboratory}
  \institution{Imperial College London}}
\renewcommand{\shortauthors}{CCS 2019 Workshop}
\begin{abstract}
 Voice-enabled interactions provide more human-like experiences in many popular IoT systems. Cloud-based speech analysis services extract useful information from voice input using speech recognition techniques. The voice signal is a rich resource that discloses several possible states of a speaker, such as emotional  state, confidence and stress levels, physical condition, age, gender, and personal traits. Service providers can build a very accurate profile of a user's demographic category, personal preferences, and may compromise privacy. To address this problem, a privacy-preserving intermediate layer between users and cloud services is proposed to sanitize the voice input. It aims to maintain utility while preserving user privacy. It achieves this by collecting real time speech data and analyzes the signal to ensure privacy protection prior to sharing of this data with services providers. Precisely, the sensitive representations are extracted from the raw signal by using transformation functions and then wrapped it via  voice conversion technology. Experimental evaluation based on emotion recognition to assess the efficacy of the proposed method shows that identification of sensitive emotional state of the speaker is reduced by $\sim$96 \%.
\end{abstract}
\maketitle
\input{Introduction.tex}
\input{Threat_Model.tex}
\input{Model_Framework.tex}
\input{Experiment.tex}
\input{Related_Work.tex}
\input{Discussion_and_Future_Work.tex}
\bibliographystyle{ACM-Reference-Format}
\balance
\bibliography{sample-base}
\end{document}

%% file: Introduction.tex
\section{Introduction}
Voice-controlled IoT services have become quite popular. Seamless interaction between the users and services are enabled through speech recognition. Many IoT devices have built-in microphones that listen for user commands, such as instructing a TV to turn on, or the coffee machine to prepare coffee. The intelligence of these services is constantly evolving so as to better understand their users' states. For instance, several online recommendation systems rely on speaker identification or emotion recognition to provide recommendations for purchases or restaurants. These suggestions may be presented based on the age of the user or the user's current mood. 

By increasing the capabilities of voice-control systems, it becomes feasible to launch new privacy and security attacks. Voice is one of the most important sources of affective data. It includes various embedded metadata such as the who, when, where and what that may be extracted from the voice signal. Sending a raw signal to a service provider's cloud for further analysis can reveal deep sensitive personal information. Alepis and Patsakis in \cite{1} presented and analysed the potential risks of voice assistants in mobile devices, showing how urgent it is to develop privacy-preserving architectures for speech analysis by extracting the distinguish features from the speech without compromising individual privacy.

Recently, computational paralinguistics has attracted the attention of researchers due to its prominent potential for practical IoT applications. It helps to understand diverse speaker states, traits, and vocal behaviours \cite{2}. One of the most popular objectives of computational paralinguistics is emotion recognition to enable naturalistic human-computer interaction. It has enhanced the quality of several cloud-based services such as for call centres, but will raise  concerns for privacy and data security. The main question here is how to analyse speech without disturbing the user's privacy in terms of removing sensitive information from the voice signal before releasing it to a third party. In this work, a privacy-preserving framework based on voice conversion is proposed to sanitize speech data. It aims to normalise a sensitive part of the speech (such as emotion state) while preserving the signal utility (speech content) before sending it to the cloud for further analysis. Firstly, the initial processing of the speech is done to calculate the features to be hidden and then used as a target to train the feature extraction model. CycleGAN architecture \cite{3} is used as the feature extractor. Then, the output features are used to re-generate the voice files using a state-of-the-art vocoder: WORLD \cite{7}.

To evaluate the trade-off between data utility and privacy, the proposed method is tested on an emotion recognition task using the RAVDESS dataset \cite{11}. The results show that the proposed solution can decrease the the accuracy of state-of-the-art paralinguistic models such as emotion recognition, while affecting the accuracy of speech recognition and speaker identification techniques only minimally. 
The code and results are available from the project page\footnote{\href{https://github.com/RanyaJumah/PP_Speech_Analysis}{https://github.com/RanyaJumah/PP-Speech-Analysis}}.

%% file: Threat_Model.tex
\section{Threat Model}
Due to resource limitations on edge devices, speech analysis is outsourced to the cloud for best performance. However, service providers aim to expand their abilities to understand the additional information about the speakers by developing models that process their voice input and detect their current conditions. They are able to collect sensitive behaviour patterns from voice input that may violate user privacy in numerous ways. They may infer a users' mental state, stress level, smoking habits, overall health conditions, indication of Parkinson's disease, sleep patterns, and levels of exercise \cite{22}. For instance, Amazon has patented technology that can analyse users' voice to determine emotions and/or mental health conditions. It allows understanding speaker commands and responses according to their feeling to provide highly targeted content \cite{29}. Therefore, speech analysis seems to play an especially important role when it comes to advertising content related to physical or emotional states. 

Emotions are a universal aspect of human speech that convey their behaviour. The physical characteristics of emotion expression in several psychiatric conditions has been investigated in \cite{21}. As a consequence of listening to users' voices and monitoring emotions, resulting critical decision-making may effect the users life, ranging from fitness trackers for well-being to suitability for recruitment. Adding these feelings or health conditions to user profiles at the service provider side will open many new privacy issues. Therefore, a proposed privacy-preserving layer is proposed and evaluated to protect the emotional privacy as a sensitive part of speech while maintaining user experience. This bridges the communication between  users and a service provider cloud, and serves as a wrapper of the emotional part of the voice input to prevent service providers from monitoring users' emotions that associated with their speech.
\newline

\begin{figure}[t!]
  \centering
  \includegraphics[width=\linewidth]{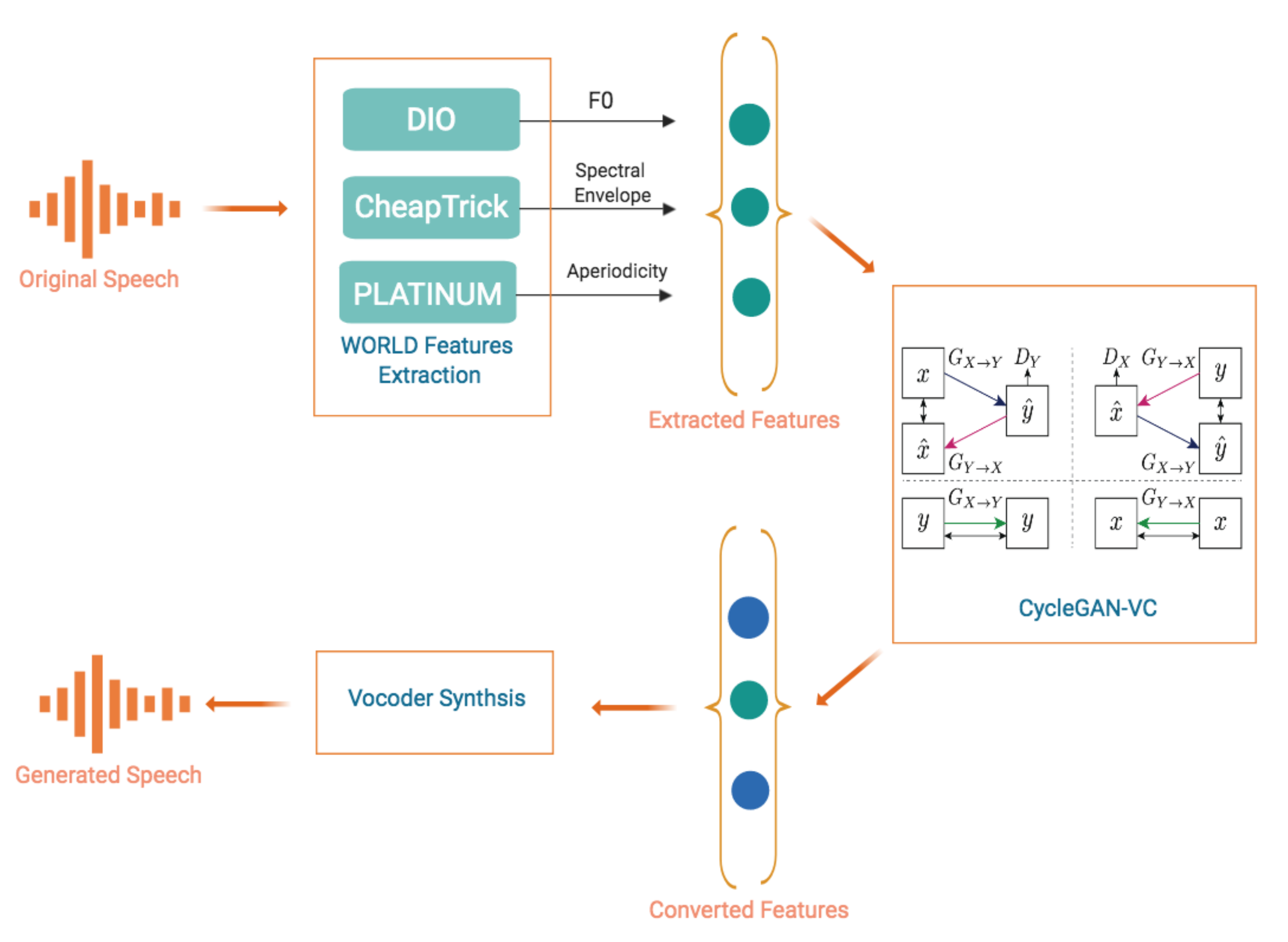}
  \caption{Block diagram of the proposed privacy-preserving framework for speech analysis}
  \Description{}
        \label{fig:framework}
\end{figure}

%% file: Model_Framework.tex
\section{Model Framework}

Generative Adversarial Networks (GANs) consist of two networks: a generator and a discriminator~\cite{4}. Generator aims to generate new data similar to the expected one, while discriminator recognizes if an input data is real or fake, as produced by the generator. CycleGAN~\cite{3} is a custom model of GANs that uses two generators and two discriminators. By considering X and Y as different domains that generators task to convert from X to Y and vice versa. Generator (G) maps from domain X to Y, and generator (F) maps from Y to X. In addition, two adversarial discriminators D (X) and D (Y), where D (X) aims to distinguish between objects in X domain and output objects from F (Y), and D (Y) aims to discriminate between (Y) and the output of G (X). 

Two objective functions are the power of CycleGAN: an adversarial loss, and a cycle consistency loss. Adversarial loss expresses the objective of the generators that attempt to fool its corresponding discriminator into being less able to distinguish its generated output from the real one. Cyclic loss calculates the loss of translating a sample from Y to X and then back again to Y. The full objective is:
\begin{equation}
Loss_{Full} = Loss_{Adv} + \lambda  Loss_{Cyc}
\end{equation}
where hyperparameter $\lambda$ controls the relative significance of the two objectives. CycleGAN is a core architecture behind the privacy-preserving framework for speech analysis that aims to learn sensitive representations in speech. Similar to the CycleGAN-VC2 model in~\cite{5}, the features that are extracted by CycleGAN can be used to transform the sensitive information in the speech with other non-sensitive data, without loss of utility for specific tasks. The proposed framework consists of the following modules, as shown in Figure~\ref{fig:framework}.

\subsection{Pre-processing}
Unsupervised learning is used to extract the representation from speech. The distinguishing signal features are extracted  by performing transformations or sampling strategies to the input data, and using the resulting outcomes as labels \cite{6}. The most effective features in speech processing are F0 counter, spectral envelope and aperiodic information. To accomplish this, WORLD is used to estimate different features of the raw speech signal using three algorithms. Firstly, fast and reliable F0 extractor is applied to find intervals of zero crossings and peaks of a waveform to estimate the fundamental frequency (F0) of the speech \cite{8}. CheapTrick is implemented to estimate the spectral envelope using F0 information \cite{9}. Finally, Definitive Decomposition Derived Dirt-Cheap (D4C)  estimates the aperiodicity of the speech signal, which is the power ratio between the speech signal and the aperiodic component of the signal, i.e. noise \cite{10}. 

\begin{figure}[t!]
  \centering
  \includegraphics[width=\linewidth]{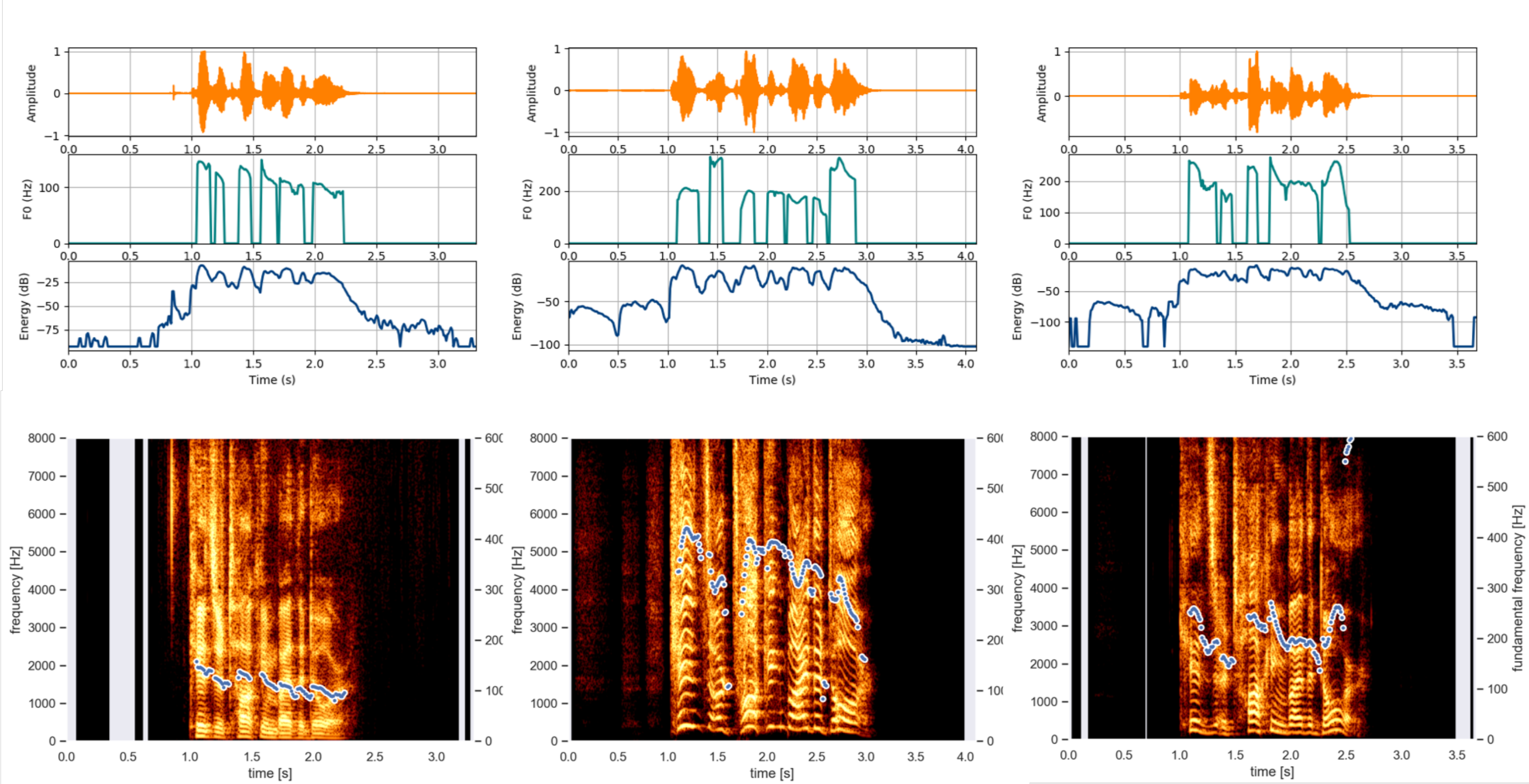}
  \caption{Acoustic features comparison between three emotional states: natural, angry, and happy respectively.}
      \label{fig:featuresComp}
\end{figure}

\begin{figure}[t!]
  \centering
  \includegraphics[width=\linewidth]{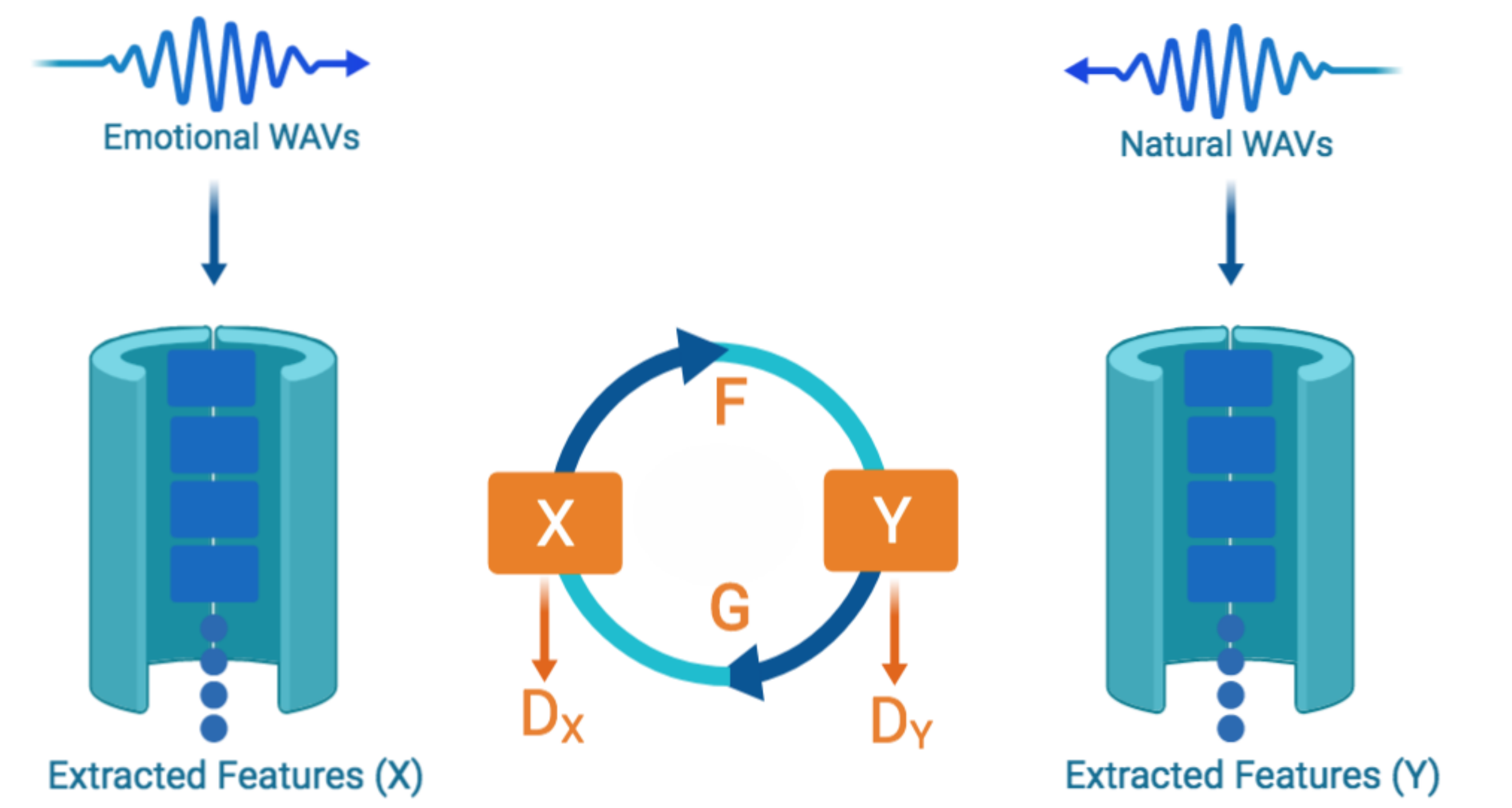}
  \caption{A basic architecture of CycleGAN
  to transform acoustic parameters of emotional utterances such that the modified speech conveys a netural utterance.}
  \Description{The 1907 Franklin Model D roadster.}
    \label{fig:cyclegan}
\end{figure}

\subsection{Conversion Process}
Leveraging non-parallel VC, the feature extraction model is trained to identify the sensitive features from the speech and convert them to non-sensitive features. For example, the prosody features are the most related in emotion recognition tasks that can be computed directly from the signal by applying transform functions. It notes that these features are completely different among emotion categories; see Figure~\ref{fig:featuresComp}. Then, a computed feature is used to train a feature extraction model to minimize the computational overhead and extract these specific features. Finally, feature-conversion is applied to hide the sensitive data, as shown in Figure~\ref{fig:cyclegan}, and re-generate the speech signal using WORLD, which has synthesis algorithm to generate high-quality synthesized speech.
\newline

%% file: Experiment.tex
\section{Experiment}
The determination of sensitive features in speech is according to the specific task for speech processing. By assuming the sensitive information to be hidden is the speaker's emotion (decrees the emotion recognition) and the desired speech processing is to analyse its content script and identify the current speaker (maintain the speech and speaker recognition), the settings for the experiment were as follows:

\subsection{Emotional Speech Dataset}
Speech audio-only files (16bit, 48kHz .wav) from the Ryerson Audio-Visual Database of Emotional Speech and Song (RAVDESS) \cite{11} were used as the dataset. It contains 1440 files: 60 recorded per actor x 24 professional actors (12 female, 12 male), vocalizing two lexically-matched statements in a neutral North American accent. Speech emotions include calm, happy, sad, angry, fearful, surprise, and disgust. While two expressive styles are used as case studies (happy and angry), the proposed method can be applied to other emotions.

\subsection{Experimental Settings}
\textbf{Pre-processing and Feature Extraction} The dataset is reconfigured by placing different emotional speech together as source and placing the natural speech as a target. Then, waves are downsampled to 16.00 kHz and acoustic parameters logarithmic fundamental frequency (log F0), spectral envelopes (SEs), and periodicities (APs) are extracted with a 5 ms frame. 

\textbf{Emotion Conversion} is another form of the voice conversion which focuses on prosody parameter transformations of the speech. In \cite{12} a data-driven emotion conversion system has proposed to output expressive speech by transforming a natural utterance to three target emotion: anger, surprise, and sadness. Moreover, Lopez et al. in~\cite{16} presented a speaking style conversion method to convert the normal utterance to Lombard speech. First, F0 is estimated and then duration and spectral conversion (vocal tract characteristics) are implemented. The key step in emotion conversion is transforming the spectral envelope of each F0 segment using CycleGAN-VC2 \cite{5}, which has been trained on the RAVDESS dataset. Spectral features are mapped using CycleGAN from utterances spoken in emotional ways to corresponding features of normal speech. Finally, the mapped features are converted to normal speech waveforms using WORLD.

\textbf{Speech Recognition} Google Cloud Speech-to-Text API is used to evaluate the speech recognition of the generated audio files. It can recognize real-time streaming using Google's machine learning technology \cite{19}.

\textbf{Speaker Recognition} is an extremely challenging task, and it requires high performance often under real-time conditions. To ensure that the proposed system is highly confident that a person is speaking and has been correctly identified, a trained model on VoxCeleb2 \cite{17} has been used \cite{18}. All audio files are converted to 16-bit streams at a 16kHz for consistency. Spectrograms are then generated of size 512 x 300 for 3 seconds of speech. Mean and variance normalisation is performed on every frequency bin of the spectrum. These spectrograms are then used as input to the CNN.

\textbf{Emotion Recognition} aims to automatically identify the affective state of the users.
An emotion classification model based on RAVDESS dataset has been used to predicts 7 emotion classes which are the following: 0 = neutral, 1 = calm, 2 = happy, 3 = sad, 4 = angry, 5 = fearful, 6 = disgust, 7 = surprised \cite{19}.
\newline

%% file: Related_Work.tex
\section{Related Work}
The studies have differed between analysing the breakthroughs on the voice-enabled systems themselves to reveal the users' privacy by analysing their communications.

\textbf{Voice-controlled Privacy}
Voice is considering as one of the unique biometric information that has been widely used in various IoT applications. Google Home, Amazon Alexa, and Apple Siri are a famous voice-based smart personal assistant.  Many privacy and security breach of voice-based systems have been reported in literature. For example, the adversary can build an acoustic model of the victims and re-generate any provided text by using that model. Spoofing the voice-based authentication system will allow the attackers to illegal access to speaker private information \cite{13}. 

\textbf{Privacy-preserving Voice Conversion}
The voice is a very significant to serve as a good index for several traits and physical characteristics. Several sensitive information has been extracted from the voice input such as emotions \cite{25} and health state \cite{23,24}. For example, the age, height, and weight of a speaker can be predicted based solely on hearing his or her voice \cite{26}. Further, the physical strength of the individuals, especially men, can be assessed based only on hearing the sound of their voice \cite{27}. Mairesse et al. \cite{28} proposed classification, regression and ranking models to learn the Big Five personality traits of a speaker. To tackle this issue, voice conversion has been used to support the preservation of user privacy by hiding the sensitive representation. The key concept of the voice conversion is how to convert the speech signal into target speaker while preserving the linguistic contents \cite{14}. In \cite{15}, VoiceMask is proposed to mitigate the security and privacy risks by concealing voiceprints and adding differential privacy. It sanitized the audio signal received from the microphone by hiding the speaker's voiceprint and then sending the perturbed speech to the voice input apps or the cloud.\cite{chung2018voxceleb2}
\newline
\begin{figure}[t!]
  \centering
  \includegraphics[width=\linewidth]{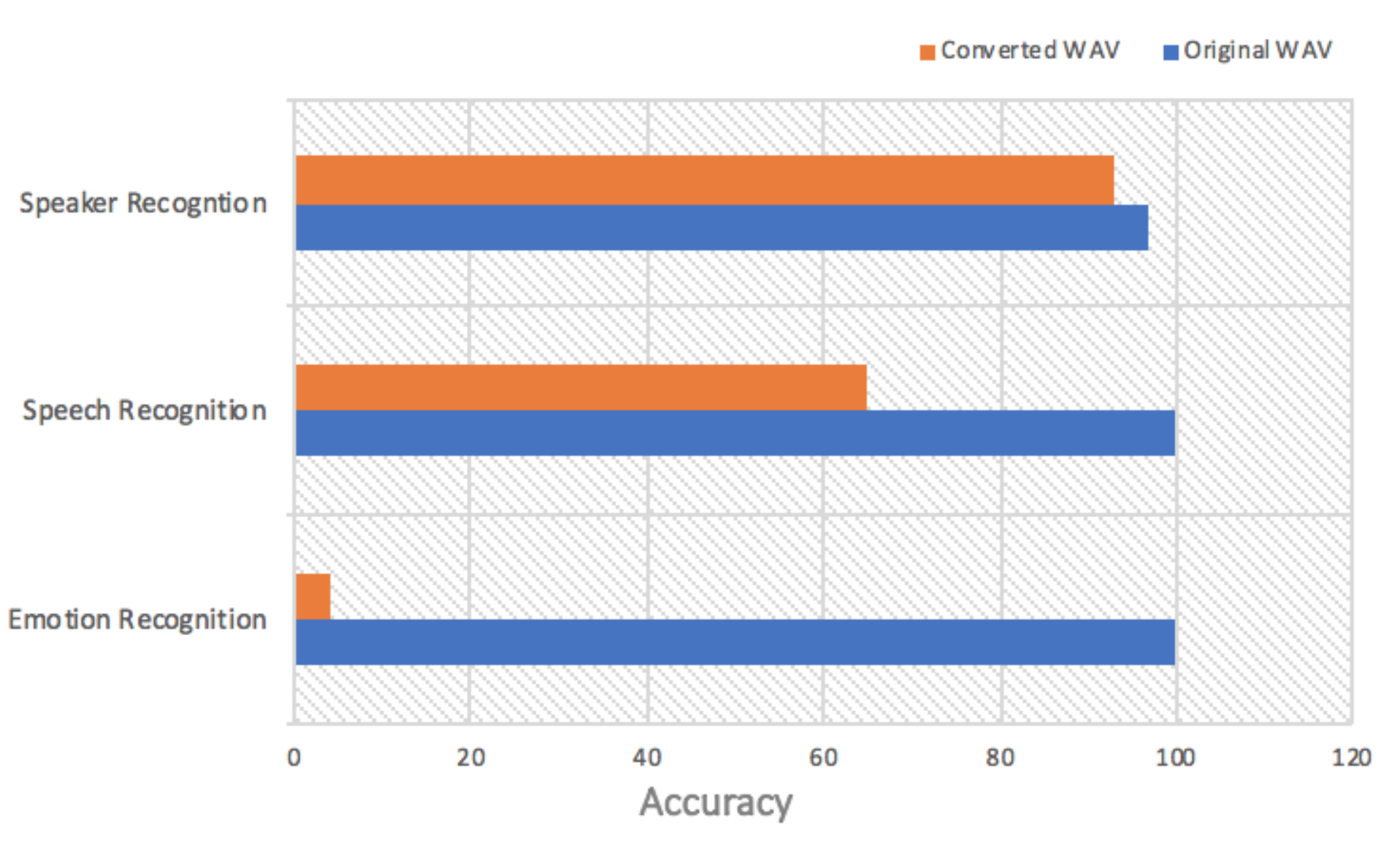}
  \caption{Accuracy comparison between three speech analysis tasks: speaker recognition, speech recognition and emotion recognition}
  \Description{}
        \label{fig:result}
\end{figure}

%% file: Discussion_and_Future_Work.tex
\section{Discussion and Future Work}
In general, the proposed framework has tested over 40 emoational recorded from RAVDASS dataset which is diffrent from the training set. These Wave files in particular style have converted to another (emotion-to-normal) via emotion conversion.  Then, a comparison of the linguistics information between the converted and the original speech signal has done to evalute the effect of the conversion process on it. The proposed method was successfully able to hide the real  emotion with drop of emotion recognition accuracy by 96 \% percent, albeit with a decrease in speech recognition accuracy where the average word error rate (WER) is 35 \%. In addition, speaker recognition performance is measured by the equal error rate (EER), which is the rate at which both acceptance and rejection errors are equal. The speaker recognition accuracy has a slight decrease of $\sim$0.12 \%. By increasing the training epoch, the results will be improved significantly. Therefore, it is shown that the proposed method can achieve the preservation of privacy in speech analysis. Figure ~\ref{fig:result} presented the comparative results between the original and generated voice signals among the three speech analysis tasks: speech recognition, speaker identification, and emotion classification.  

Voice-based systems continue to enhance user experience. Therefore, voice anonymisation method for voice input is suggested to trade-off between the signal utility and speaker privacy. The challenge is how to sanitize the speech without degrading the speech recognition accuracy. The evaluation results show the effectiveness of the proposed method in terms of projecting away sensitive representations (emotion) while preserving the speech quality. Further experiments will be conducted for speech analysis and achieving adaptive speech recognition while preserving speaker privacy. Additionally, to further strengthen the user privacy we will include filtering speech content to prevent similar outcomes using other techniques, such as sentiment analysis.
\newline